# Biological Nonlocality and the Mind-Brain Interaction Problem: Comments on a New Empirical Approach

Fred Thaheld
99 Cable Circle #20, Folsom, Calif. 95630 USA

*The more perfect a nature is the fewer means it requires for its operation.*
                                                                Aristotle

**Abstract**

Up to now we have been faced with an age old fundamental dilemma posed by the mind-brain interaction problem, i.e., how is it that the mind which is subjective and immaterial, can interact with the brain which is objective and material? Analysis of recent experiments appears to indicate that quantum mechanics may have a role to play in the resolution of the mind-brain interaction problem in the form of biological entanglement and nonlocality. In addition this analysis, when coupled with ongoing and proposed experiments, may help us to simultaneously resolve related issues such as whether mental events can initiate neural events, the transference of conscious subjective experience, the measurement problem and the binding problem.

*E-mail address:* fthaheld@directcon.net

# 1. Introduction

Over 50 years ago Popper (1953) proposed an interactionistic mind-brain theory in the following fashion: "There is no reason (except a mistaken physical determinism) why mental states and physical states should not interact. (The old argument that things so different could not interact was based on a theory of causation which has long been superseded)". This was followed by additional related material (Popper, 1955; 1972: chapter 6;1973;1978; Popper & Eccles, 1977).

Later he discussed several critical aspects of the mind-brain problem, and proposed a hypothesis dealing with this situation in force-field terms (Popper et al, 1993).

"I think the special problem of the body-mind issue is: apparently there exist forces with a certain amount of autonomy from the, let us say, physiological entity, with which these forces are related and with which they interact. In other words, we commonly think of forces as mere appendices to matter. Apparently, forces which are related to biochemical substances can obtain a certain autonomy and independence from these sheer substantial forces".

"And what we really have to study is: How can these forces, which are set up in the brain, continue themselves, so to speak, and continue to have a kind of identity which is even able to initiate in its turn biochemical processes in the brain? That seems to me to be the body-mind problem".

"I wish to propose here as a hypothesis that the complicated electro-magnetic wave fields which, as we know, are part of the physiology of our brains, represent the unconscious parts of our minds, and that the conscious mind - our conscious mental intensities, our conscious experiences - are capable of interacting with these unconscious physical force fields, especially when problems *need* to be solved that *need* what we call "attention". This admittedly vague working hypothesis appears to me as a small yet significant progress within a so far hopelessly difficult part of physiology".

In their interpretation of Popper's interactionastic hypothesis, Lindahl & Århem (1994) advance the idea that there are two levels of interaction: the first between a certain spatio-temporal pattern of action potentials and a specific electromagnetic field (the relations 1 and 2 as per their Fig. 1); the other between the electromagnetic field/the unconscious, and the unconscious mind (the relations 3 and 4). Also, that the distinction between the conscious mind and the brain is seen as a division into what is subjective and what is objective, and not as an ontological distinction between something immaterial and something material. And, that this interpretation confronts us with the problem of explaining how objective processes can interact with subjective processes.

As was pointed out (Beck, 1996) the problem is that the mind needs a physical force if it acts on the brain but, that Popper has not supplied us with a definition of the mind as a force field and, there is apparently a missing link between mind and brain. He further goes on to state that, "The occurrence of probability fields in quantum tunneling has led John Eccles and the author (Beck & Eccles, 1992), on the basis of the quantum trigger model for exocytosis, to postulate a mind-brain interaction by selection of quantum amplitudes analogous to a measuring process". And, "That seems to us to be the crucial point behind why quantum processes in neuronal activities are essential for an interactionistic mind-brain theory avoiding difficulties with well-established concepts (strong conservation laws) of the physical world" (Beck, 1996). He further feels that, "The need to introduce quantum processes into brain dynamics, on the other hand, opens up exciting perspectives for new experimental endeavors in neuroscience".

Lindahl & Århem (1996) in their response to Beck (1996) reject his idea of a missing link and feel that interactionism does not necessarily imply an immaterial-material dualism which might violate the strong conservation laws of physics. They do not find it necessary to postulate an intermediate field of probability amplitudes. They do agree with Beck that quantum processes may have an explanatory value in the development of a viable interactionistic mind-brain theory and, that concentrating on the possible role of quantum processes in brain dynamics opens up exciting perspectives.

**2. Nonlocality and entanglement**

Einstein, Podolsky and Rosen (1935) wrote a paper in which they claimed that if quantum mechanics were a complete model of reality, then nonlocal correlations between particles had to exist. This has become well known as the EPR paradox or effect and, is based upon the assumption that correlations exist between particles which have interacted in the past and then separated, which interaction has resulted in the particle states becoming what Schrödinger (1935) termed *entangled*. It is helpful to quote him on this subject in the interest of clarification in this difficult interdisciplinary area.

"When two systems, of which we know the states by their respective representatives, enter into temporary physical interaction due to known forces between them, and when after a time of mutual influence the systems separate again, then they can no longer be described in the same way as before, viz. by endowing each of them with a representative of its own. I would not call that *one* but rather *the* characteristic trait of quantum mechanics, the one that forces its entire departure from classical lines of thought. By the interaction the two representatives (or wave-functions) have become *entangled*".

To account for correlations between these particles, Bell (1964) considered theories which invoke common properties of the pair. With the addition of a reasonable locality assumption, he showed that such classical-looking theories are constrained by certain inequalities that are not always obeyed by quantum mechanical predictions and, proposed certain experiments to test for locality assumptions. Years later an experiment (Aspect et al, 1982) showed that nonlocal influences do indeed exist after these particles (in this instance photons) interact and, that a polarization measurement on one of them, using variable polarizers, instantaneously resulted in the other photon possessing the same degree of polarization. This does not imply any transference of energy or a signal to accomplish this feat, as this would be in conflict with special relativity. These space-like separated photons are connected or *entangled* by virtue of having interacted in the past and, even though no energy or signal passed between them, there appeared to be a superluminal transfer of *information* or *influence* (Stapp, 1977).

It should be further noted that one can test the explicit quantum nature of systems via the use of EPR nonlocality since, as per Feynman (1982), this nonlocality cannot be duplicated by a classical system. Additional experiments on *entangled* particle states (photons) have further verified the nonlocal nature of quantum mechanics (Tittel et al, 1998; Thew et al, 2002) out to a distance of several km. It should also be stated at this juncture that the problem of understanding the consequences of the EPR nonlocality effect and entanglement is a controversial issue in itself (Laloe, 2001) where just nonliving particles are concerned at the microscopic level.

At the present time according to Josephson, it appears that we may have to differentiate between two different kinds of nonlocality, one related to quantum mechanical nonlocality and the other to what he has termed 'biological utilization of quantum nonlocality' (Josephson & Pallikari, 1991). In addition he Josephson, 1988) has explored the issue of whether significant differences do exist between the living organism and the type of system studied so successfully in the physics laboratory. And, whether one can deal with living organisms in quantum mechanical terms, with the same degree of rigor as is normal for non-living particle systems. In the interest of brevity I have chosen to reduce this term to *biological nonlocality*.

**3. A review of past and present experiments indicative of biological nonlocality**

The first experiments which appeared to indicate that a nonlocal relationship might exist between the brains of individuals were originally performed by Grinberg-Zylberbaum (1982; Grinberg-Zylberbaum and Ramos, 1987; Grinberg-Zylberbaum et al, 1994). This indicated that a visual evoked potential (VEP) elicited in the brain of one subject by unpatterned photostimulation, gave rise to an evoked potential possessing a similar brain wave morphology or, what they termed a *transferred potential* in the brain of another nonstimulated subject located several meters away. They considered this EPR-style nonlocality, since both of them were in Faraday chambers which ruled out most electromagnetic and all neural influences. It is of interest to note here that while there was no indication of transference of conscious subjective experience between any of the subject-pairs, the appearance of a *transferred potential* was accompanied by the participants *feeling* that their interaction had been successfully completed.

Recently a group from Bastyr University and the University of Washington proposed to replicate the Grinberg-Zylberbaum experiments, with emphasis on the nature of the evoked potential signal, the baseline characteristics of EEG coherence and the remote nonlocal influence in human neurophysiology (Richards & Standish, 2000). They subsequently received a 2 year grant from the National Institutes of Health (NIH) to pursue this objective and, after completion of their research have achieved some preliminary results which appear to lend credence to the concept of remote nonlocal influence (Standish, 2003; Standish et al, 2004). Their research revealed that 9 out of 20 subject pairs, selected on the basis that they either knew each other, were related or had interacted in the past in some other fashion such as through meditation, had statistically significant ($p<.01$) effects with the nonstimulated subject (receiver) synchronized to the stimulated subject's (sender) VEP (Richards et al, 2002). I.e., there were changes in the receiver's alpha rhythm that corresponded to the on/off conditions of the sender's checkerboard pattern visual photostimulation, which they referred to as evidence of 'neural energy transfer'. Although the subject pairs were separated by 30 feet in two separate rooms, these were not Faraday chambers and so the results cannot be said to definitely represent evidence of remote nonlocal influence. Also, I question their use of the term 'neural energy transfer' as this is not compatible with their stated replication emphasis regarding 'remote nonlocal influence'.

In addition, a second group at the University of Freiburg has successfully replicated the Grinberg-Zylberbaum experiments in a more robust fashion, having observed this phenomenon of a *transferred potential* in 11 out of 14 subject pairs to date while the subject pairs were in Faraday chambers (Seiter et al, 2002; Walach et al, 2003; Wackermann et al, 2003). They split up their subject pairs into two groups consisting of experimental and control subjects. Experimental subjects were invited as 'empathically bonded and connected pairs', while subjects in the control group were totally unrelated, did not meditate or interact with each other in any fashion and were told nothing about the nature of the experiment. Their interesting finding was that *both* groups showed significant changes in their EEG variance in at least one channel ($p=0.01$ for the experimental group and $p=0.09$ for the control group). This could mean that *entanglement* and remote nonlocal influence are generic phenomena which are applicable to all living entities and, in the case of humans, naturally arising based upon the discovery of similar mitochondrial DNA in blood samples which have been taken from people around the world, implying prior entanglement in some fashion.

Recent experiments conducted at the University of Freiburg (Wackermann et al, 2003), continue to show a high co-incidence of variations of the brain electrical activity in the nonstimulated subjects with brain electrical responses of the stimulated subjects. I.e., there are detectable stimulus-related 'correlations' between brain activities of two separated subjects. They feel that we are facing a phenomenon which is neither easy to dismiss as a methodical failure or a technical artifact nor, understood as to its nature. And, that no biophysical mechanism is presently known that could be responsible for the observed EEG correlations.

They also feel that the results should not be interpreted as a successful replication of the *transferred potential* (as was supposedly seen in prior experiments) in that they did not see any VEP-like wave-forms in the averaged EEG of the non stimulated subjects. I.e., it is not an exact *transferred copy* of the VEP in the stimulated subject. Thus, the term *transferred potential* as I have used it in this paper can cover a wide range from VEP-like wave-forms to simultaneous brain wave-form events in the nonstimulated subject's VEP but, not necessarily exact VEP-like wave-forms.

It is of interest to note here that recent research indicates that an extremely minute aspect of biological quantum nonlocality has been observed in the coherence of induced magnetic dipoles involved in muscle contraction involving single actin filaments (Matsuno, 2001; 2002; Hatori et al, 2001). The way that they distinguished between classical and quantum coherence is the robustness of the coherence. If it is classical the coherence will be completely up to the boundary conditions provided externally. The classical coherence is vulnerable even to the slightest changes in the boundary conditions. If it is quantum the coherence can maintain its robustness even in the presence of disturbances to some extent. The observed magnetization was robust against thermal fluctuations, which should effectively deal with objections raised by some physicists that biological quantum coherence and superposition is impossible due to ambient thermal considerations (Tegmark, 2000). Compelling arguments have recently been advanced in rebuttal to Tegmark and supporting quantum coherence (Hagan et al, 2002).

## 4. Analysis of previous and proposed experiments predicated upon the concept of biological nonlocality

At the Tucson 2000 Toward a Science of Consciousness conference I proposed that it would be possible to conduct experiments in the immediate future in which we could deal simultaneously with three of the major problems within the field of consciousness studies (Thaheld, 2000a) from the standpoint of what I have termed *biological nonlocality*. First, whether there are nonlocal correlations between two human mind-brains. Second, whether one can transfer conscious subjective experience from one individual to another. And third, whether *mental events* can influence, control and initiate cerebral events or what is known as the *reverse direction* problem (Libet, 1994), not only in the brain of one individual but, more importantly, *between* the brains of two subjects. I believe that these three entities may be intimately interrelated for the following reasons:

1. If both subjects are in separate Faraday chambers (as has been discussed in Sec. 3), this effectively rules out most electromagnetic and all neural energy transfer mechanisms, leaving *biological nonlocality* as one of the leading mechanisms.

2. Furthermore, if my proposal to utilize special patterned photostimulation (figures, symbols, pictures, etc.) rather than the normal checkerboard photostimulation is successful (Thaheld, 2000a; 2001; 2002) and, the nonstimulated subject is able to give us a conscious subjective experience report on 'seeing' some modicum of this pattern which he *received* from the stimulated subject and, which is coincidental with the appearance of a *transferred potential* on his EEG, this would reveal that conscious subjective

-10-

experience had been transferred from a photostimulated individual to a nonstimulated subject.

3. Since we know that most electromagnetic and no neural energy could have passed through 2 Faraday chambers, plus the intervening layers of resistance present in both brains (dura mater, skull, scalp), we are left with only *mental events* from the stimulated subject as having been able to initiate, control and influence cerebral events in the nonstimulated subject, or what is known as the *reverse direction.* This argument makes use of existing concepts of quantum mechanics and does not rely upon any new and unknown force fields, such as the 'conscious mental field' (CMF) proposed by Libet (1994), which have no experimental underpinnings at the present time.

At the Skovde, Sweden 2001 Toward A Science of Consciousness conference, I proposed additional experiments involving *biological nonlocality*, dealing with the measurement problem, or whether consciousness collapses the wave-function, and the binding problem, or how it is that the brain can fuse together the many disparate features of a complex perception (Penrose, 1994;Thaheld, 2000b; 2001).

As regards the measurement problem, I reasoned that if the simultaneous EEG of the nonstimulated subject, representing the *transferred potential*, revealed a similar brain wave morphology or simultaneous brain wave-form events related to the stimulated subject's VEP (but, not necessarily exact VEP-like wave-forms), that we might be looking at the actual collapse of the wave-function in a living system, since there did not appear to be any back-reaction resulting from the *transferred potential* of the nonstimulated subject back to the stimulated subject, which you might expect in a strictly classical situation. And, that if we are able to achieve a transference of conscious subjective experience, this might provide the nonstimulated subject with visual evidence of the actual collapse of his wave-function, in addition to what he might see on his own EEG.

In effect this becomes somewhat like the original Aspect experiment with photon twins except in this instance we are performing a measurement on the stimulated subject with the patterned photostimulation and eliciting a similar response from the nonstimulated subject, which we can readily discern from the *transferred potential.* The only differences being that Aspect was dealing with *entanglement* at the microscopic individual photon level (nonliving entity), whereas here we are dealing with what appears to be macroscopic *entanglement* (living entity). The other major difference is that each time after he made a measurement on the photons, there was a collapse of the wave-function and, he had to prepare a new pair of photons for the next measurement. In the present instance after we make a measurement, it appears that it may be possible through the utilization of *biological nonlocality* for the macroscopic *entangled* living system to constantly maintain or regenerate these *entangled states* for varying periods of time. This analysis seems to be borne out by the Hz rate of photostimulation over a period of minutes which has been achieved in the Freiburg experiments.

Regarding the binding problem, the VEP (which results from a large number of neurons being activated as a result of patterned photostimulation from a large number of photons falling upon the retina of the stimulated subject) is simultaneously accompanied by a *mental event* or *force* (information or influence transfer). This *mental event* gives rise to a *transferred potential* in the nonstimulated subject, again from a large number of neurons but miraculously, without *one photon* falling upon the retina! Not only was the mind or a *mental force* at work but, it appears that a large number of neurons in both the stimulated and nonstimulated subjects' brains had to be *entangled* in a quantum mechanical fashion both between themselves and within their individual brains in order for the *transferred potential* to appear. The brain of the stimulated subject seems to fuse together a very few disparate features of a simple perception, in this instance the patterned photostimulation, and this is conveyed by the *mental event* to the nonstimulated subject where it, in turn, activates a large number of *entangled* neurons, resulting in a *transferred potential* as is revealed on the EEG.

These experimental proposals should also allow us to examine the question of the Neural Correlates of Consciousness (NCC), which is defined as a specific system in the brain whose activity correlates directly with states of conscious experience. One then has to ask if the *transferred potential* in the nonstimulated subject is the equivalent of the NCC, especially since the *transferred potential* results from a simultaneous *mental event* in the stimulated subject elicited by specific patterned photostimulation, which is further revealed and accompanied by a VEP. The nonstimulated subject receives this *mental event* from the stimulated subject, which is directly converted back into an analogous neural event, or a *transferred potential*, which the nonstimulated subject perceives as a NCC or a conscious subjective experience. This sequence of events has led me to postulate the existence of the NCC's equal and dual partner, which I have named the *Mental Correlates of Consciousness* (MCC).

-13-

**5. Transcranial Magnetic Stimulation and the transference of conscious subjective experience.**

At this same conference I also proposed the use of Transcranial Magnetic Stimulation (TMS) in addition to specific patterned photostimulation, in an attempt to achieve a more strikingly visible and repeatable transference of conscious subjective experience (Thaheld, 2001). This is a non-invasive technique of directly stimulating the human cortex using a pulsed magnetic field, without any discomfort to the subject and requiring no direct contact with the scalp (Barker et al, 1985). Stimulation is assumed to be due to the current induced in the tissue by the rapid time-varying field and, results in a very prominent and distinctive evoked potential in the EEG record. This selective stimulation of different cortical areas at variable rates from 1-40 Hz can also be correlated with a measured external response, such as an evoked motor potential in a particular muscle group such as the hand (thenar) or leg (tibialis muscle). Subjects also report phosphenes or, the sensation of light in darkness, as a result of TMS over the occiput.

TMS could be administered to a stimulated subject, giving rise to either an evoked motor potential in a particular muscle group or the appearance of phosphenes. We would then want to see if these same evoked motor potentials or phosphenes are elicited in the nonstimulated subject in a similar manner and, in addition to the *transferred potentials.*

**6. Concluding remarks**

In conclusion, I would like to make the following comments:

(i) Research which has been going on for several years now at various universities continues to show that something very unusual is being successfully replicated involving *entanglement* and *biological nonlocality* between human subjects.

(ii) By utilizing the mind-brains of two subjects simultaneously, we will now be better able to explore what is going on in just the mind-brain of one subject.

(iii) Since signals or energy cannot be transferred in nonlocal fashion due to the constraints imposed by special relativity, could such substituted terms as *information* or *influence transfer*, embraced within the concept of a *mental event*, be the equivalent of Popper's *force field*? I.e., Popper's *force field* may already be encompassed by quantum mechanics.

(iv) It would now appear that the *mental event* simultaneously appearing with the VEP from a stimulated subject, initiated biochemical processes (neural events) in the brain of a nonstimulated subject, which was revealed in the appearance of a *transferred potential*. To paraphrase Popper (1993), "these forces which were set up in the brain of one subject, continued themselves, so to speak, and continued to have a kind of identity which was able to initiate in its turn biochemical processes in the brain of a second subject" This sequence of events would appear to require both a classical and nonlocal neuronal depolarization initiation process.

**Acknowledgment**

I wish to thank the editor, reviewers and Thesa von Hohenastenberg-Wigandt for their comments and suggestions.

.